\documentclass[letterpaper, 10 pt, conference]{ieeeconf}  

\IEEEoverridecommandlockouts                              

\usepackage[pdftex]{graphics} 
\usepackage{epsfig} 
\usepackage{mathptmx} 
\usepackage{times} 
\usepackage{amsmath} 
\usepackage{amssymb}  

\usepackage{xcolor}
\usepackage{multirow}
\usepackage{graphicx}
\usepackage{booktabs}

\usepackage{multicol}
\usepackage[bookmarks=true]{hyperref}
\usepackage{subcaption}
\usepackage[capitalize]{cleveref}
\crefname{section}{Sec.}{Secs.}
\Crefname{section}{Section}{Sections}
\Crefname{table}{Table}{Tables}
\crefname{table}{Tab.}{Tabs.}
\newcommand{\B}{\bfseries}

\title{\LARGE \bf
Sim2Real Transfer for Audio-Visual Navigation with Frequency-Adaptive Acoustic Field Prediction
}

\author{Changan Chen$^{1*}$, Jordi Ramos$^{1*}$, Anshul Tomar$^{1*}$, Kristen Grauman$^{1,2}$
\thanks{* indicates equal contribution, sorted in alphabetical order}
\thanks{$^{1}$University of Texas at Austin, $^{2}$FAIR, Meta AI}%
}

\begin{document}

\maketitle
\thispagestyle{empty}
\pagestyle{empty}

\begin{abstract}
Sim2real transfer has received increasing attention lately due to its success in transferring robotic policies learned in simulation to the real world. While significant progress has been made in transferring vision-based navigation policies, the current sim2real strategy for audio-visual navigation remains limited to basic data augmentation. Sound differs from light in that it spans across much wider frequencies and thus requires a different solution for sim2real. To understand how the acoustic sim2real gap varies with frequencies, we first define a novel acoustic field prediction (AFP) task that predicts the local sound pressure field. We then train frequency-specific AFP models in simulation and measure the prediction errors on collected real data. We propose a frequency-adaptive strategy that intelligently selects the best frequency band for prediction based on both the measured prior and the energy distribution of the received audio, which improves the generalization on real data. Coupled with waypoint navigation, we show the navigation policy not only improves navigation performance in simulation but also transfers successfully to real robots. This work demonstrates the potential of building autonomous agents that can see, hear, and act entirely from simulation, and transferring them to the real world.

\end{abstract}
\vspace{-0.05in}
\section{Introduction}
Navigation is an essential ability of autonomous robots, allowing them to move around in the environment and execute tasks such as delivery, search and rescue. Sometimes, the robot also needs to hear the environment and navigate to find where the sound comes from, e.g., when someone is asking for help in a house, or when the fire alarm goes off.

Navigation has been extensively studied in the robotics community and has been traditionally approached with Simultaneous Localization and Mapping (SLAM)~\cite{slam_survey} with Lidar sensors, which is limited in semantic reasoning. Recent research has increasingly focused on vision-centric navigation, where robots rely primarily on visual sensors to perceive their environment, showing significant success in photorealistic real-scanned settings~\cite{deitke22retrospectives}. Various tasks have been proposed, such as PointGoal navigation~\cite{zhu17targetdriven,gupta17cognitive,savva19habitat}, ObjectGoal navigation ~\cite{chaplot20object,batra20objectnav}, or visual exploration~\cite{ramakrishnan22poni,chen21learning,chaplot20learning}. Other work further expands the sensory suite to include hearing. In particular, the AudioGoal task~\cite{chen20soundspaces,gan20look} requires an agent to navigate to a sounding target (e.g., a ringing phone) using audio for directional and distance cues while using vision to avoid obstacles in the unmapped environment.

With the success of these learning-based navigation systems in simulation, efforts have been made to transfer learned policies to the real world by addressing the sim2real gap~\cite{anderson20simtoreal,peng18simtoreal,Kadian2020}. 
However, recent work~\cite{gao23sonicverse} on audio-visual navigation with sim2real transfer through data augmentation, which fails to account for the influence the range of frequencies have on the sim2real gap. In this work, we systematically evaluate the acoustic gap and propose a solution to bridge it.

\begin{figure}
    \centering
    \includegraphics[width=0.65\linewidth]{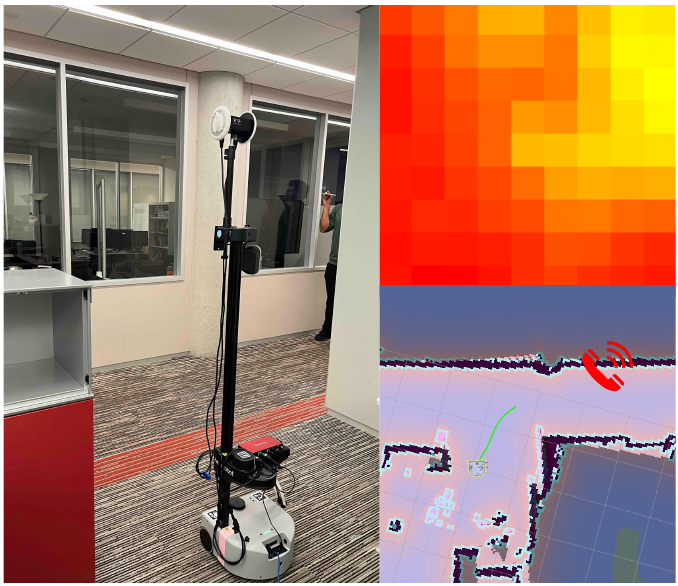}
    \vspace{-0.05in}
    \caption{Our robot predicts an acoustic field with a frequency-adaptive model and navigates to locate the sound source.}
    \label{fig:concept}
    \vspace{-0.25in}
\end{figure}

State-of-the-art approaches in audio-visual navigation rely on reinforcement learning to train the navigation policy end-to-end~\cite{chen20soundspaces,chen21semantic}, which not only poses challenges for interpretation but also struggles to generalize to the real world due to various sim2real gaps. 
Recent advancements in visual navigation have shown successful sim2real transfer with hierarchical models~\cite{anderson20simtoreal,ramakrishnan22poni}, which consist of a high-level path planner and a low-level motion planner. This hierarchical design helps mitigate some of the low-level physical discrepancies encountered during transfer.
However, existing hierarchical models have not yet attempted to address sim2real for audio-visual navigation.

Inspired by these methods, we design a modular approach to ease the transfer from simulation to the real world. To achieve this, we confront a key question: what is the proper high-level planning task that can survive sim2real transfer for audio-visual navigation? 
To this end, we propose a novel prediction task: \emph{acoustic field prediction}---predicting the local sound pressure field around the agent. The gradient of this field reflects the direction of the sound. 
Measuring acoustic fields is expensive in the real world since it requires simultaneously capturing the sound pressure of all points in the field due to the dynamic nature of sound. 
However, computing acoustic fields is free in simulation.  We first build an audio-visual model as the acoustic field predictor (AFP) and curate a large-scale acoustic field dataset on SoundSpaces 2.0~\cite{chen22soundspaces2}, the state-of-the-art audio-visual simulation platform. We show that this approach outperforms existing methods on the Continuous AudioGoal navigation benchmark.

After validating the proposed approach in simulation, we then investigate where acoustic discrepancy arises. It is known that ray-tracing-based acoustic simulation algorithms introduce more errors with lower frequencies due to wave effects~\cite{savioja2015overview}.
Given this observation, we focus on evaluating how sim2real errors vary with frequencies. We first collect real acoustic field data with the source sound being white noise, whose audio energy uniformly spans across all frequencies. We then train acoustic field prediction models that only take the sub-frequency band of the input audio and test it on the real white noise data. By computing the errors across multiple samples, we show that the errors do not strictly go down as the frequency goes higher, and using the best frequency band yields errors smaller than using all frequencies for the white noise sound.

However, simply taking the best frequency band does not work for all sounds since different sounds have varying spectral distributions. 
To address this issue and make the model aware of the spectral difference, we propose a novel frequency-adaptive prediction strategy, which selects the optimal frequency sub-band based on measured errors and the spectral distribution of the received audio. To validate this approach, we collect more acoustic field data with various sounds and show that the frequency-adaptive model leads to the lowest error on the real data compared to other strategies.

Lastly, we build a robot platform that equips the Hello Robot with a 3Dio binaural microphone and then deploy our trained policy on this robot. We show that our robot can successfully navigate to various sounds with our trained frequency-adaptive acoustic field prediction model. See \cref{fig:concept} and Supp. video.

In summary, we propose a novel acoustic field prediction approach that learns to navigate without interaction with the environment. This approach improves the SOTA methods on the challenging Continuous AudioGoal navigation benchmark~\cite{chen22soundspaces2}. We perform a systematic evaluation of the sim2real challenge and propose a frequency-adaptive strategy as the treatment for sim2real. We show this strategy works on both collected real data as well as on our robot platform.  
To the best of our knowledge, this is the first work to investigate and propose a principled solution to the sim2real transfer problem for audio-visual navigation.
\vspace{-0.05in}
\section{Related Work}

\vspace{-0.05in}
\subsection{Embodied Navigation}
\label{sec:embodied_visual_navigation}
To navigate autonomously, traditionally a robot builds a map via 3D reconstruction  (i.e., SLAM) and then plans a path using the map~\cite{FuentesPacheco2012VisualSL}. 
Recent works have developed navigation policies that make navigation decisions in a previously unmapped environment from egocentric observations directly without relying on mapping~\cite{gupta17cognitive,batra20objectnav,wijmans20ddppo}.
Some recent efforts have developed audio-visual simulation platforms~\cite{chen20soundspaces,chen22soundspaces2,gan21threedworld} that enable embodied agents to both see and hear.
In the AudioGoal navigation task~\cite{chen20soundspaces,chen21semantic}, the agent must navigate to the source of sound in an unknown environment. The state-of-the-art audio-visual navigation models train policies with reinforcement learning (RL), requiring millions of samples. 
Inspired by recent work in learning potential functions for interaction-free navigation~\cite{ramakrishnan22poni}, we propose to predict acoustic fields for interaction-free audio-visual navigation.

\vspace{-0.05in}
\subsection{Sound Localization in Robotics}
\label{sec:sound_localization_in_robotics}
In robotics, microphone arrays are often used for sound source localization~\cite{nakadai1999sound,rascon2017localization,nakadai2000active}. 
Past studies fuse audio-visual cues for surveillance~\cite{wu2009surveillance,qin2006learning}, speech recognition~\cite{yoshida2009automatic}, human robot interaction~\cite{alameda2015vision,viciana2014audio}, and robotic manipulation tasks~\cite{romano2013ros}. These solutions typically depend on analytical solutions for computing the direction of sound, whose performance deteriorates under strong reverberation and noise~\cite{beamforming_reverb}. Recent works propose learning-based sound localization~\cite{rao22learning,jenrungrot2020cone}, which however require collecting real data for training. In this paper, we show that we can transfer models trained in simulation directly to the real world.

\vspace{-0.05in}
\subsection{Sim2real Transfer}
\label{sec:sim2real_transfer}
Benefiting from recent large-scale datasets of real-world 3D scans~\cite{xia18gibson,chang17matterport3d} and supporting simulators ~\cite{savva19habitat,ai2thor}, recent works have shown success in enabling embodied agents in simulation.
Transferring the model trained in simulation to the real world is thus of great interest. The mostly widely used approaches are domain randomization~\cite{Tobin2017,Juliano2019sim2real}, 
system identification~\cite{Kadian2020,Kristinn_identification}, 
and transfer learning and domain adaptation~\cite{fuzhen2020survey}. 
Most sim2real research studies transferring a policy from simulation to the real world based on visual input. Recent work~\cite{gao23sonicverse} does sim2real transfer for audio-visual navigation by applying data augmentation empirically, which does not account for the effect of frequencies on sim2real.In this work, we systematically evaluate the sim2real gap by collecting data and identifying the spectral discrepancy.
\vspace{-0.05in}
\section{Approach}

\vspace{-0.05in}
\subsection{SoundSpaces Platform and Audio-Visual Navigation}
We first introduce the simulator that we are using. SoundSpaces 2.0~\cite{chen22soundspaces2} is a state-of-the-art audio-visual simulation platform that produces highly realistic audio and visual rendering for arbitrary camera and microphone locations in 3D scans of real-world environments~\cite{chang17matterport3d,straub19replica,xia18gibson}. It accounts for all major real-world acoustics phenomena: direct sounds, early specular/diffuse reflections, reverberation, spatialization, and materials and air absorption.

In audio-visual navigation, the goal is to navigate to a sounding object in an unknown environment by seeing and hearing. The location of the sound source is not known and needs to be inferred from the received audio. At every step, the agent needs to sample an action from \{MOVE\_FORWARD, TURN\_LEFT, TURN\_RIGHT, STOP\}. If the agent issues the STOP action within a 1m radius of the goal location, the episode is considered successful.

There are different instantiations of the audio-visual navigation task, each with its goal specification. For example, AudioGoal~\cite{chen20soundspaces} requires navigating to a static target in a discretized environment, while Dynamic AudioGoal~\cite{dynamic_av_nav} requires navigating to a moving sound source. 
We target the Continuous AudioGoal navigation benchmark introduced in SoundSpaces 2.0~\cite{chen22soundspaces2}, which 
generalizes the state space to be the continuous environment.

\vspace{-0.1in}
\subsection{A Modular Design for Sim2real Transfer}
\vspace{-0.05in}
Transferring a navigation policy trained in simulation
to the real world is not trivial due to many domain gaps between the simulation environment and the real world, which include the visual discrepancy, the physical dynamics discrepancy, the robot actuation discrepancy and---specifically in this task---the acoustic discrepancy. 

We focus on investigating the acoustic discrepancy and to bridge other domain gaps (e.g., visuals and physics), we take a hierarchical approach that disentangles navigation into high-level path planning and low-level motion planning. This has a few benefits: 1) disentangling the policy makes it possible to utilize existing SLAM algorithms on the real robot to abstract away domain gaps other than the audio. 2) disentangling the policy makes the intermediate output more interpretable and easier to debug 3) specifically in this work, posing the high-level planning as a supervised prediction task makes it easier to measure the sim2real difference because we can evaluate the performance by collecting real measurements without repeatedly running robots.

The key challenge here is to formulate the proper waypoint prediction task 
that could survive the sim2real transfer. One existing approach~\cite{gan20look} predicts the exact location of the audio goal directly, which is however an ill-posed problem since the environment geometry is unknown. For example, when the audio goal is in another room, the received audio reveals the direction to the door rather than the exact direction of the goal.
Instead of predicting the global audio goal location, we propose to predict the local acoustic field (sound pressure field) centered around the agent, which not only better captures the direction of the sound but also is more predictable from the visual observation of the environment.

\begin{figure}[t]
    \centering
    \includegraphics[width=0.9\linewidth]{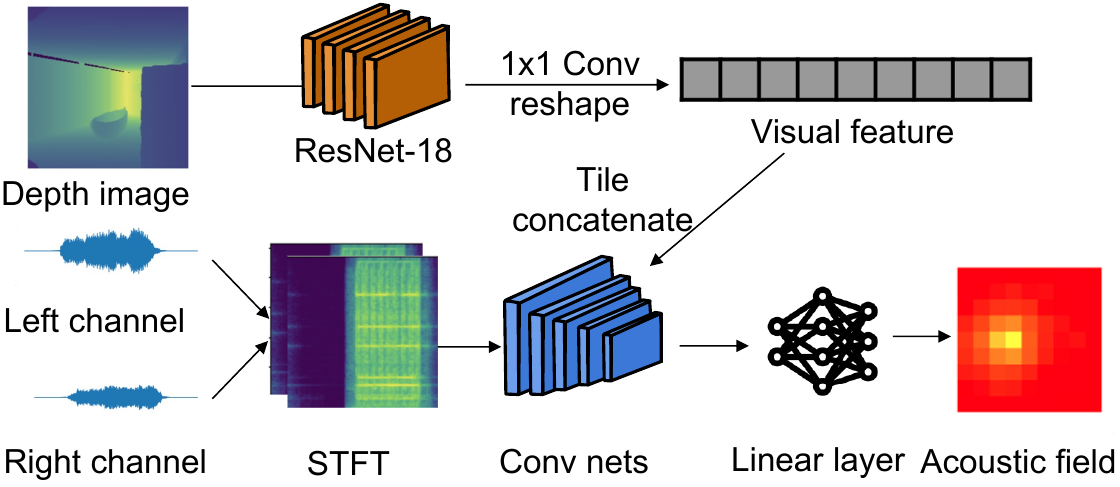}
    \vspace{-0.1in}
    \caption{Acoustic field prediction model. 
    The model first extracts audio and visual features, and then tiles and concatenates features channel-wise to predict the acoustic field.
    }
    \label{fig:acoustic_field_predictor}
    \vspace{-0.2in}
\end{figure}

Defining the waypoint prediction task however does not address the audio discrepancy directly.
SoundSpaces 2.0 renders the audio propagation as a function of the geometry of the environment, the material properties, and source/receiver locations based on a bidirectional ray-tracing algorithm~\cite{cao2016interactive}. While it produces realistic audio renderings, there remains some difference between how sound propagates in the real world vs in simulation. It is known that ray-tracing-based algorithms yield worse performance with lower frequencies due to wave effects~\cite{savioja2015overview}. This implies the model needs to be aware of this spectral difference for sim2real.
Thus we introduce a frequency-adaptive prediction strategy to help the model better transfer to the real world in \cref{sec:frequency_adaptive}.

\vspace{-0.05in}
\subsection{Acoustic Field Prediction}
\vspace{-0.05in}
We define acoustic field prediction as predicting the top-down sound pressure field of $L\times L$ centered at the agent, given the egocentric depth image of $128\times128$ pixels and a one-second binaural audio.

To tackle the acoustic field prediction problem, we first present a model that uses both audio and visual observations (see \cref{fig:acoustic_field_predictor}). The motivation behind using the visual sensor is that the visual observation of the surrounding environment can be useful in inferring the geometry of the environment, which affects the acoustic field. For example, walls often act as the boundary of the acoustic field.
More specifically, we use a pre-trained ResNet~\cite{he16deep} to extract features from the input image and reshape it with a $1\times 1$ Conv layer into a 1d vector of size $512$. For the input binaural audio, we first process the waveforms with Short-Time Fourier Transform (STFT) to convert the time-domain signal into the frequency domain. We then use a 2D Conv net to encode the features and then tile and concatenate with the visual feature. Lastly, we feed the final output through one linear layer and reshape the prediction into the size of the target acoustic field $L\times L$.

\vspace{-0.1in}
\subsection{Hierarchical Navigation}
\begin{figure}[t]
    \centering
    \includegraphics[width=\linewidth,keepaspectratio]{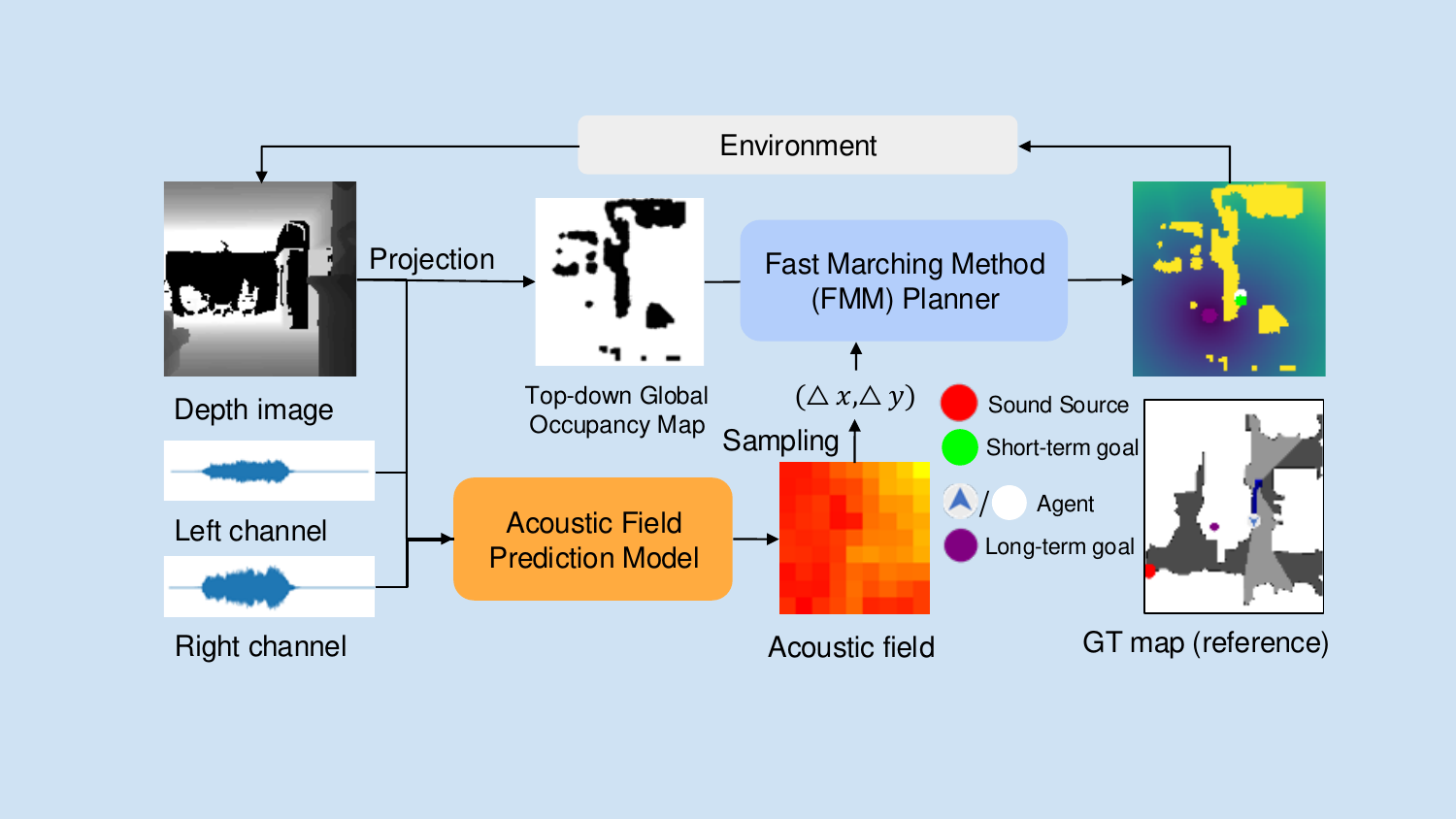}
    \vspace{-0.25in}
    \caption{Navigation pipeline. The model first predicts the acoustic field, samples the peak as the long-term goal, and navigates toward the goal with a path planner.
    }
    \label{fig:hierarchical_navigation}
    \vspace{-0.2in}
\end{figure}

\vspace{-0.05in}
With this acoustic field prediction model, we then construct a hierarchical navigation pipeline (see \cref{fig:hierarchical_navigation}) to perform audio-visual navigation, which executes the following steps: 1) sampling a long-term goal; 2) navigating to the long-term goal; and 3) making the stopping decision. 

\subsubsection{Sampling a Long-Term Goal}
At each time step, the agent predicts the acoustic field based on audio-visual inputs and then identifies the maximum value of the field. We set the peak location as the long-term goal either when there is no existing long-term goal or the new peak value surpasses the value of the existing long-term goal since as the agent gets closer to the goal, the sound usually gets louder.

\subsubsection{Navigating to the Long-Term Goal}
After sampling the long-term goal, for path planning, we use the Fast Marching Method (FMM)~\cite{sethian1999fast} to determine the best route to the goal in simulation.
FMM takes the occupancy map that is built on the fly, the agent's current position, and the long-term goal as inputs. The occupancy map is computed by calculating the point cloud observed at each timestep using the depth camera. Next, FMM calculates the distance between each navigable point in the map to the long-term goal. The algorithm then selects the adjacent point on the map with the lowest value as its short-term goal and the agent then moves towards that point. 
When the long-term goal is sampled at a non-navigable location, we use breadth-first search (BFS) to find the closest available point to navigate to.

\subsubsection{Stopping Criteria}
The stopping condition is evaluated each time after the agent reaches a long-term goal or the closest navigable point to the long-term goal. When the agent samples a new long-term goal, if the peak value of the predicted acoustic field is at the center of the field, the agent issues the stop action.

\vspace{-0.05in}
\subsection{Frequency-adaptive Prediction}
\label{sec:frequency_adaptive}
Existing audio-visual navigation models use all frequencies in the input audio. However, sound spans across a wide range of frequencies, and due to imperfect geometric simulation techniques, the acoustic gap varies as a function of frequencies.
Models trained with all frequencies assuming them equally reliable would have lower performance when deployed on a real robot.

\begin{figure}[t]
    \centering
    \includegraphics[width=0.65\linewidth]{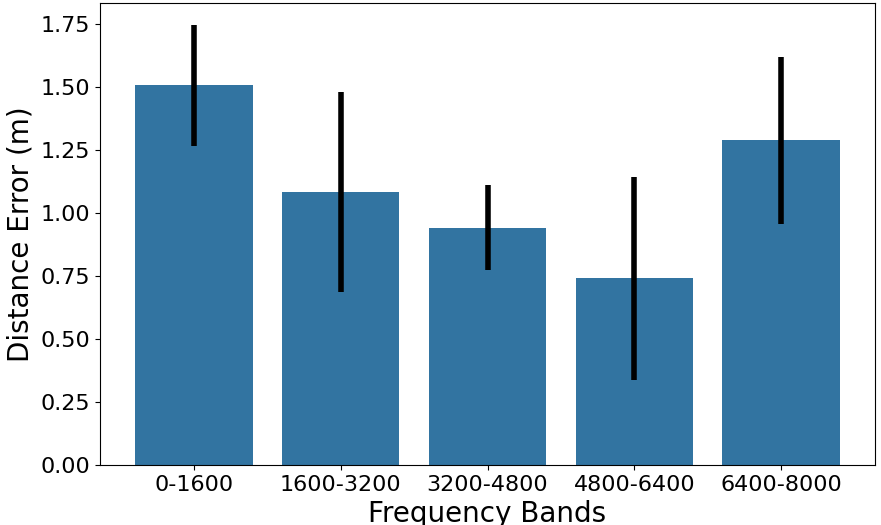}
    \vspace{-0.05in}
    \caption{Sim2real error as a function of frequencies. We report the mean and standard deviation of distance errors between the predicted and the ground truth peak locations.}
    \label{fig:spectral_discrepancy}
    \vspace{-0.25in}
\end{figure}

Given this observation, we first systematically examine how the gap changes as a function of the frequency. The idea is simple: with a given frequency band $[F_1, F_2]$, we first train an acoustic field prediction (AFP) model in simulation using only that band, then test it on real-world data of the same sound and same band, and calculate the prediction error. 
We equally divide all frequencies into $N$ subbands ($N=5$ 
based on hyperparameter tuning), and we show the distribution of errors over the frequency bands in \cref{fig:spectral_discrepancy}, where error is defined as the distance between the predicted max and ground truth goal location.
As expected, the lower frequencies tend to yield larger prediction errors.  However, the error does not monotonically decrease as frequency increases likely due to simulation errors. We also trained a model that uses all frequencies, which has a distance error of $0.86 m$, underperforming the best frequency band.

With this measurement, the most intuitive idea would be just to take the frequency band that has the least sim2real error and train a model with that band. However, this will not work for real-world scenarios where some sounds span across many frequency bands while others only occupy a very narrow range of frequencies. To take that into account, we propose a frequency-adaptive prediction strategy that uses the best frequency band based on both the measured error and the energy distribution of the received audio.

Assume we divide all frequencies linearly into $N$ bands. Given a received audio $A_r$, we first convert it into the frequency domain and divide it into these $N$ bands. Based on the measured errors, we have a weighting function that assigns weights to these bands based on their sim2real errors: 
\vspace{-0.15in}
\begin{equation}
    p(i) = (\frac{1}{e_i})^\alpha, i \in [1, ..., N],
\vspace{-0.05in}
\end{equation}
where $e_i$ is the error in \cref{fig:spectral_discrepancy}. For each subband $i$ of the input, we then compute another weight based on the energy of the band normalized with respect to the highest energy: 
\vspace{-0.05in}
\begin{equation}
    q(i) = (\frac{r_i}{r_m})^\beta, i \in [1, ..., N],
\vspace{-0.05in}
\end{equation}
where $r_i$ is the energy of that band and $r_m = \max_i r_i$. We basically assign higher weights to frequency bands that have more energy. Lastly, we take the product of these weights: 
\vspace{-0.05in}
\begin{equation}\label{eq:weighting}
    w(i) = p(i)\times q(i), i \in [1, ..., N]
\vspace{-0.05in}
\end{equation}
We take band $i$ with the highest $w(i)$ to produce the final prediction. Both $\alpha$ and $\beta$ are hyper-parameters, and we perform a grid search to find the best values on validation.

Intuitively, what the weighting function does in \cref{eq:weighting} is: if the input sound has a fairly equal distribution of energies over all subbands, it will take the best band from $p(i)$ that has the lowest sim2real error. If the input sound has a very skewed energy distribution, it will prioritize taking the band where the audio has the most energy. In this way, we factor into both the measured difference and the spectral distribution of individual sounds.

\vspace{-0.05in}
\subsection{Implementation and Training Details}
For the size of the acoustic field, we set $L$ to $9$ with a grid resolution of $0.5m$, i.e., $4.5m\times 4.5m$ centered around the agent based on our ablations. $\alpha$ and $\beta$ are set to $5$ and $0.8$ respectively based on the validation performance.

We train the predictor with Mean Squared Error (MSE) loss till convergence. For optimization, we use the Adam optimizer~\cite{kingma15adam} with a learning rate set of $0.001$.
\vspace{-0.05in}
\section{Data Curation}\label{acoustic_field_dataset}
\vspace{-0.05in}
Due to the expense of measuring real acoustic field data, we choose to utilize simulation to collect large-scale training data.
We also collect real data for measuring the sim2real gap and validating our frequency-adaptive prediction model.

\vspace{-0.05in}
\subsection{SoundSpaces Acoustic Field Dataset}
SoundSpaces 2.0 supports computing the impulse response $I(s, r)$ between the source location $s$ and the receiver location $r$ as a function of the 3D environment but does not have direct API support for rendering the acoustic field. To compute the field, given a $(s,r)$ pair, we first sample a grid centered at the receiver location of size $L\times L$, and for each grid point $p$, we compute $I(s, p)$ which results in $L^2$ number of RIRs per receiver location. However, these RIRs are represented in the form of waveforms instead of single numbers. To best represent the sound pressure at each single point, we take the maximum amplitude of the waveform. 

For sampling the source/receiver locations and environments, we utilize the existing audio-visual navigation episodes~\cite{chen20soundspaces}, which provides configurations of the environment and source/receiver locations. This dataset uses scenes from the Matterport3D dataset~\cite{chang17matterport3d}, which contain scans of real-world environments such as apartments, offices, and even churches. We sample 500 episodes per environment for the 57 training environments in the navigation dataset. We also perform a similar operation to curate the validation and test set. In total, we collect 1.1M/52K/52K samples for train/val/test. Along with these acoustic fields, we also render the RGB-D images at the corresponding locations. See examples in \cref{fig:acoustic_field_prediction}.

\vspace{-0.05in}
\subsection{Real Measurements Collection}
\vspace{-0.05in}
To measure the sim2real error, we collect real audio measurements to evaluate the trained model's performance. For that, we use a 3Dio microphone to capture the binaural audio with a smartphone serving as the speaker output. We align the real-world parameters closely with those in our simulator, such as the height of the speaker and receiver. Since the simulator employs a mono receiver, the two-channel audio data we gather is transformed into mono format by averaging the amplitude values across both channels. 
This process is repeated for ten distinct speaker positions (8 different directions w.r.t the agent and two data points for when the speaker is near the agent).
We also downsample the acoustic field resolution from $9\times 9$ to $3\times 3$ so that we can collect more data in more environments.

For the source of the sounds, we use two types of sounds: white noise and normal sounds. To compute the sim2real errors in \cref{fig:spectral_discrepancy}, it is important for the sound to have uniform distribution across all frequencies, and we use white noise for that. For evaluating the final frequency-adaptive acoustic-field prediction model, we choose 7 unheard sounds that have varying spectral distributions
and play them as the source. For each sound, we collect 10 data points. We split them equally into validation and test for hyperparameter searching.

\vspace{-0.2in}
\section{Robot Platform}
\label{sec:real_robot_navigation}
\vspace{-0.05in}
To deploy our sim2real policy on a real robot, we build our audio-visual robot by equipping a HelloRobot with a 3Dio binaural microphone as shown in \cref{fig:concept}. We use Focusrite Scarlett Solo as the audio interface to amplify the audio signals from the binaural microphone.

To start the navigation, we first sample the current audio from the microphone and predict the long-term goal from the acoustic field. We then pass this goal to the robot and use HelloRobot's navigation stack to move the robot towards the goal.
Once the robot reaches the long-term goal, it comes to a complete stop for a second to sample the audio again. This process is repeated until the predicted goal location is in the center of the acoustic field. If the sampled long-term goal is in an inaccessible region, we have a time limit of 5 seconds after which the robot stops and samples a new goal.

\begin{table}[t]
    \caption{Results of the AudioGoal navigation experiment.  Our model strongly outperforms existing methods.
    }\label{tab:audiogoal_nav}
    \vspace{-0.05in}
    \centering
    \begin{tabular}{c|c|c|c}
    \toprule
         & SR $\uparrow$ &  SPL $\uparrow$ & Soft SPL $\uparrow$ \\ 
    \midrule
        Random &0.01& 0.07&0.12\\
        DDPPO~\cite{wijmans20ddppo} & 0.82& 0.63 & 0.66 \\
        Direction Follower~\cite{chen21learning} & 0.67& 0.50& 0.48\\
        Beamforming~\cite{OpenSoundscape}  &0.02 &0.01 &0.24 \\
        Gan et al.~\cite{gan20look} & 0.63& 0.53 & 0.68  \\
    \midrule
        AFP w/ predicting max & 0.54 & 0.34 & 0.38  \\
        AFP w/o vision & 0.84 & 0.71 & 0.72 \\
        AFP (Ours) &\B 0.91 &\B 0.76 &\B 0.75  \\ 
    \bottomrule
    \end{tabular}
    \vspace{-0.2in}
\end{table}

\vspace{-0.05in}
\section{Experiments}

\vspace{-0.05in}
\subsection{Results on Continuous AudioGoal Navigation Benchmark}
\label{sec:exp_audiogoal_nav}
We first demonstrate the effectiveness of our navigation system on the challenging Continuous AudioGoal navigation benchmark~\cite{chen22soundspaces2}, where the agent moves in a continuous unseen environment to find the location of a ringing telephone sound.
For metrics, we use the common Success Rate (SR), success weighted by inverse path length (SPL), and soft SPL. An episode is considered successful when the agent issues the stop action within 1 meter of the goal. SPL~\cite{batra20objectnav} is defined as $\mathrm{SPL_i} = S_i \cdot l_i / \mathrm{max}(p_i,l_i)$, where $i$ denotes the index of the episode, $S=1$ when the episode is successful and $S=0$ otherwise, $l$ denotes the length of the shortest path between the agent and the audio goal, and $p$ denotes the length of the actual path taken by the agent in the episode. Soft SPL is a variation of SPL where $S_i = 1$ for all $i$.

We compare with the following models:
\textbf{DDPPO}~\cite{wijmans20ddppo}: 
an end-to-end reinforcement learning policy trained with distributed proximal policy optimization.
\textbf{Direction Follower}~\cite{chen21learning}: this model predicts the direction of the audio goal and navigates with the same waypoint planner. We stop the agent automatically when it is within a 1-meter radius of the goal.
\textbf{Gan et al.}~\cite{gan20look}: this model predicts the (x,y) location of the audio source and navigates using a waypoint planner. 
The agent stops whenever it reaches its predicted location or the closest navigable point.
\textbf{Beamforming}~\cite{OpenSoundscape}: classical beamforming method that calculates the direction of arrival of the sound and navigates with the same waypoint planner.

To further justify our model design choice, we also compare with the following ablations of our own model.
\textbf{AFP w/ predicting max}: this model does not predict the whole acoustic field. Instead, it predicts a single point that represents the highest point of the local acoustic field. 
\textbf{AFP w/ audio-only}: this model only takes in the audio input, which tests whether the full model uses the visual information when predicting the acoustic field. 

Results are shown in \cref{tab:audiogoal_nav}. Our model strongly outperforms all baselines and ablations. Compared to DDPPO, our model is more efficient due to its hierarchical nature since the DDPPO model often gets stuck with obstacles and corners. Direction Follower and Gan et al. predict the goal direction/location directly, which is however ill-posed when goals are in some other room at a distance from the robot. As a result, their navigation performance is also pretty poor. For the Beamforming baseline, similar to ours, it also predicts the local direction of arrival of the sound; however, since it is not robust to reverberation and noise, it performs 
poorly. Lastly, the two ablations perform comparably to baselines but also underperform the full model, showing it is both beneficial to predict the full acoustic field and leverage visual sensors to understand the environment.

\begin{figure}[t]
    \centering
    \includegraphics[width=0.8\linewidth]{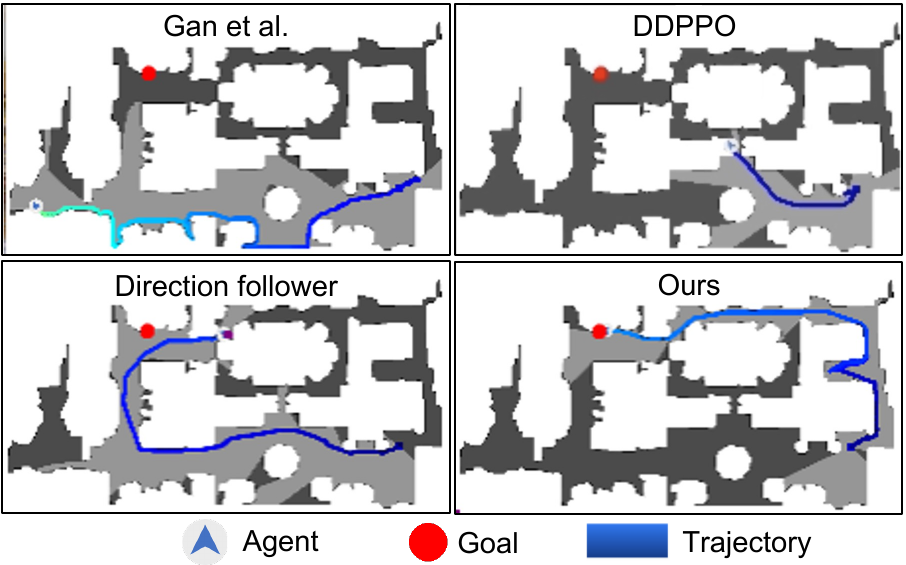}
    \vspace{-0.1in}
    \caption{Navigation trajectory comparison. Our model successfully navigates to the source while other baselines fail due to  either getting stuck or navigating in the wrong direction.
    }\vspace{-0.1in}
    \label{fig:navigation_traj}
\end{figure}

\begin{figure}[t]
    \centering
    \includegraphics[width=0.8\linewidth]{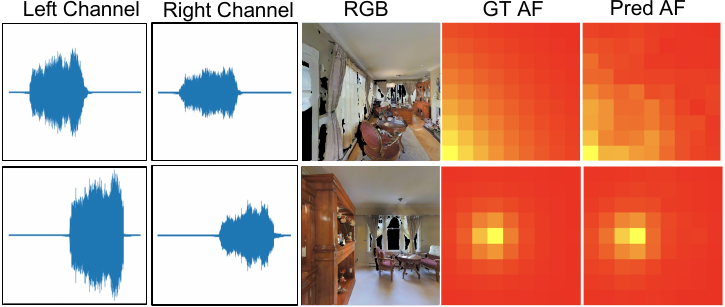}
    \vspace{-0.05in}
    \caption{Visualization of acoustic field prediction within the same episode. Top row: when the robot is still far from the goal. Bottom row: when the robot is right next to the goal. Our model predicts accurately in both cases.}
    \label{fig:acoustic_field_prediction}
    \vspace{-0.2in}
\end{figure}

We show the comparison of trajectories with these baselines in the same episode in \cref{fig:navigation_traj}, where our model is more efficient in reaching the goal.
We also visualize the acoustic fields of both the ground truth and prediction in \cref{fig:acoustic_field_prediction}. Initially, the model predicts high values at the corner (in the direction of the goal), and as the agent gets close to the goal, it predicts high values at the center of the field. 

\vspace{-0.05in}
\subsection{Experiment 2: Acoustic Field Prediction on Real Data}
Here we evaluate our frequency-adaptive acoustic field prediction (FA-AFP) model on the collected real acoustic field data. We compare our method to the random baseline and ablations of our approach. We consider three ablations: ``All-freq AFP" uses all frequencies for prediction. ``Best-freq AFP" uses the best frequency band shown in \cref{fig:spectral_discrepancy} and ``Highest-energy AFP" uses the band where the received audio has the highest energy. We measure the performance of different prediction errors with the angle and distance of the predicted max location on the acoustic field. We train our models on 73 sounds and test on 7 unheard sounds.

The results are shown in \cref{tab:afp_real}. We show that compared to the random prediction, our All-freq AFP model reduces the prediction error drastically. If we always use the best frequency for prediction, it helps lower the angle error a bit but not the distance error. Using the frequency band with the highest energy brings down the prediction error more. Our frequency-adaptive prediction model (FA-AFP) improves the performance even further, showing the importance of intelligently selecting a frequency band for prediction.

\begin{figure}
    \centering
    \includegraphics[width=0.8\linewidth]{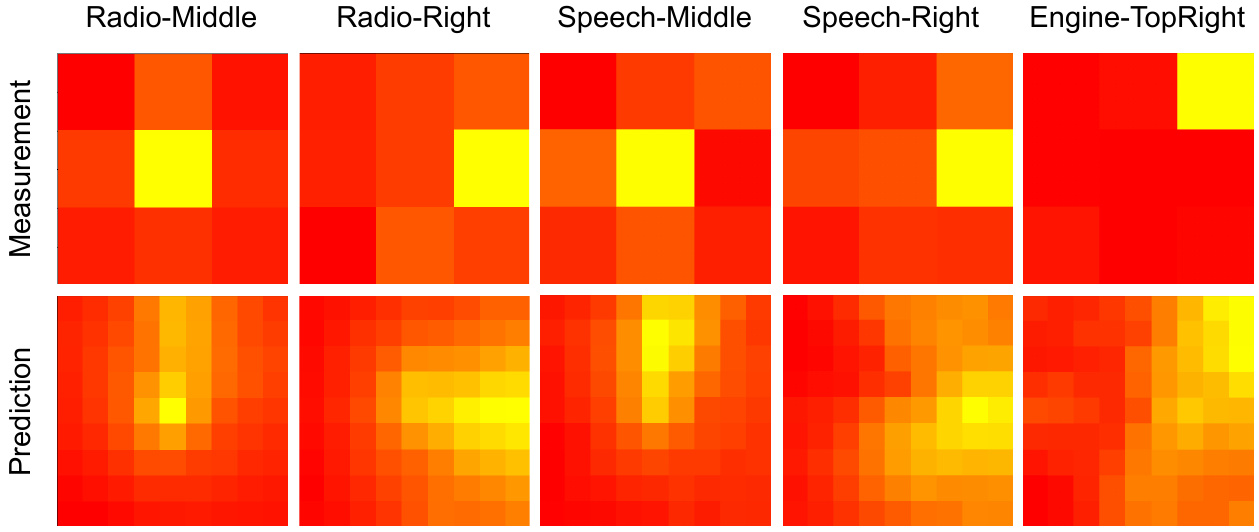}
    \vspace{-0.05in}
    \caption{Acoustic field predictions on real data. The real data is measured with a lower resolution. We show the prediction and measurement for multiple sounds and directions.
    Our model predicts all of these cases accurately. 
    }\label{fig:afp_vis_real}
    \vspace{-0.2in}
\end{figure}

In \cref{fig:afp_vis_real}, we show examples of the collected acoustic field and the predicted acoustic field for multiple directions and sounds. Note that the acoustic field is only sampled at a $3\times3$ grid centered at the robot to reduce the cost of collection. Our predictions are consistently accurate across examples.

\begin{table}[t]
    \caption{Results for testing on real acoustic field data.}
        \vspace{-0.05in}
    \label{tab:afp_real}
    \centering
    \begin{tabular}{c|c|c}
    \toprule
         &  Angle $\downarrow$ & Distance $\downarrow$\\
    \midrule
    Random    &  1.57  & 1.45\\
    All-freq AFP & 0.22  & 0.74 \\
    Best-freq AFP & 0.20  & 0.74 \\
    Highest-energy AFP & 0.04  & 0.70 \\
    FA-AFP (Ours) &\B  0.04 &\B 0.63 \\
    \bottomrule
    \end{tabular}
    \vspace{-0.1in}
\end{table}

\vspace{-0.05in}
\subsection{Experiment 3: Real Robot Navigation}

Finally, to validate the whole navigation pipeline, we deploy our navigation policy on the real robot platform (described in \cref{sec:real_robot_navigation}).
When deploying on the real robot, one thing that differs from the previously collected real data is that the robot also makes some low-frequency noise while running. To address this issue, we collect recordings of the robot noise and perform data augmentation by adding the noise to the received sound during training to improve the model performance.

We conduct 20 navigation examples with various source/receiver distances and directions
and show that our robot can navigate the sounding object with a 75\% success rate.
See the supplementary video for both success and failure cases. We also tried to deploy the best-performing baseline DDPPO, which however failed all the test scenarios, which is likely due to the significant physical sim2real gap since that model trains with RL end-to-end. We show one navigation step example in \cref{fig:concept}, where the model predicts the acoustic field correctly.
\vspace{-0.05in}
\section{Conclusion}
\label{conclusion}
\vspace{-0.05in}

We systematically evaluate the sim2real acoustic gap with a proposed acoustic field prediction task. We further design a frequency-adaptive strategy to mitigate sim2real errors. We validate our model on both the Continuous AudioGoal navigation benchmark and collected real measurements. Lastly, we build a robot platform and show that we can successfully transfer the policy to the real robot.

While this work represents an exciting first step, it also has some limitations.  First, the validation and test data were collected within the same environment, leaving the generalization to novel acoustic environments yet to be explored. Second, we assume the sound sources to be static, which may not hold in all cases, calling for new solutions to address dynamic objects.

\noindent Acknowledgements: UT Austin is supported in part by NSF CCRI and IFML NSF AI Institute. KG is paid as a research scientist by Meta.

\bibliographystyle{IEEEtran}
\bibliography{changan_general,changan_specific}

\end{document}